\documentclass[aps,twocolumn,prl,showpacs,preprintnumbers,amsmath,amssymb]{revtex4}

\def\Co{CeCoIn$_5$}

\def\CoCd{CeCo(In$_{\rm 1-x}$Cd$_{\rm x}$)$_5$}
\def\CoHg{CeCo(In$_{\rm 1-x}$Hg$_{\rm x}$)$_5$}

\usepackage{graphicx}
\usepackage{dcolumn}
\usepackage{bm}
\usepackage{txfonts}
\usepackage{ams}

\begin{document}
\bibliographystyle{E:/Work/etc/prsty}
\preprint{APS/123-QED}

\title{Anisotropic effect of Cd and Hg doping on Pauli limited superconductor CeCoIn$_5$}
\author{Y. Tokiwa$^1$}
\author{R. Movshovich$^1$}
\author{F. Ronning$^1$}
\author{E. D. Bauer$^1$}
\author{P. Papin$^1$}
\author{A. D. Bianchi$^2$}
\author{J. F. Rauscher$^3$}
\author{S. M. Kauzlarich$^3$}
\author{Z. Fisk$^4$}
\affiliation{$^1$Los Alamos National Laboratory, Los Alamos, New Mexico 87545, USA}

\affiliation{$^2$D\'{e}partement de physique Montr\'{e}al, Universit\'{e} de Montr\'{e}al, QC, H3C 3J7 Canada }

\affiliation{$^3$Department of Chemistry, University of California, Davis, California 95616, USA}

\affiliation{$^4$Department of Physics, University of California at Irvine, Irvine, California 92697, USA}

\date{\today}

\begin{abstract}
We investigated the effect of Cd and Hg doping on the first order superconducting (SC) transition and the high field-low temperature SC state of
CeCoIn$_5$ by measuring the specific heat of CeCo(In$_{\rm 1-x}$Cd$_{\rm x}$)$_5$ with x=0.0011, 0.0022 and 0.0033 and \CoHg\ with x=0.00016,
0.00032, and 0.00048 at temperatures down to 0.1\,K and fields up to 14\,T. Cd substitution rapidly suppresses the cross-over temperature $T_{\rm
0}$, where the superconducting transition changes from second to first order, to $T$=0\,K with x=0.0022 for $H\parallel$ [100], while it remains
roughly constant up to x=0.0033 for $H\parallel$ [001]. The associated anomaly of the proposed FFLO state in Hg-doped samples is washed out by
x=0.00048, while remaining at the same temperature, indicating high sensitivity of that state to impurities. We interpret these results as supporting
the non-magnetic, possibly FFLO, origin of the high field - low temperature state in \Co.

\end{abstract}

\maketitle
 In most type-II superconductors (SC) the superconducting upper critical field $H_{c2}$  is largely determined by the orbital
limiting field $H_{c2}^0$, when the opposite forces that magnetic field applies to the electrons with opposite momenta break up the Cooper pair. The
Zeeman energy of electron spins' in magnetic field also influences $H_{c2}$, and in some cases, such as two-dimensional (2D) SC (e.g. organics) with
magnetic field applied within the 2D planes, or heavy fermion compounds, with large $H_{c2}^0$, can be the dominant mechanism of suppression of SC.
In the normal state, electron spins align preferentially with magnetic field, lowering their total energy, and leading to the temperature-independent
Pauli susceptibility. In spin-singlet superconductors, the superconducting pairs are formed by electrons with opposite spins, which therefore can not
take advantage of the Zeeman energy. When Zeeman energy in the normal state is greater than the superconducting condensate energy, superconductivity
is destroyed. This effect leads to an upper bound on $H_{c2}$, called the Pauli limiting field $H_{\rm P}$~\cite{clogston:prl-62}. The relative
strength of the orbital and Pauli limiting is reflected by the Maki parameter $\alpha=\sqrt{2}H_{c2}^0/H_P$. When orbital limiting is neglected
($\alpha=\infty$), and Pauli limiting is the only effect leading to suppression of superconductivity, the SC transition is expected to become first
order below a cross over temperature $T_0 = 0.56T_c$~\cite{maki:ptp-64,SARMAG:theiue}. A large Zeeman energy in the normal state should also lead to
a peculiar SC state. Fulde and Ferrell~\cite{fulde-ferrell:pr-64} and Larkin and Ovchinnikov~\cite{larkin-ovchinnikov:jetp-64} (FFLO) predicted that
a spatially modulated SC state, that takes advantage of the electron's Zeeman energy, will be stabilized in high fields for Pauli limited
superconductors. Gruenberg and Gunther~\cite{gruenberg:prl-66} later put a lower bound on the Maki parameter ($\alpha>1.8$) for the existence of the
FFLO state.

The discovery of heavy-fermion unconventional superconductivity in CeCoIn$_5$~\cite{petrovic:jpcm-01,movshovich:prl-01} has lead to numerous
investigations of its unusual properties. \Co\ is in a strong Pauli limiting regime, with a Maki parameter $\alpha$  anisotropic with respect to the
magnetic field, ranging between 3.5 ($H \parallel$ [001]) and 4.5 ($H \perp$ [100]). CeCoIn$_5$ provided the first example of a first order SC
transition in a bulk superconductor, in accord with the above mentioned theoretical expectations, revealed in the specific heat and magnetization
anomalies at high fields~\cite{bianchi:prl-02,tayama:prb-02}. The specific heat anomaly, associated with the SC transition, sharpens and becomes more
symmetric as the field is increased, and magnetization as a function of field shows a step at $T_c$ with hysteresis. In high magnetic fields within
the basal plane of the tetragonal crystal structure of \Co, specific heat shows an anomaly inside the SC state, reflecting the formation of an
additional phase in the high-field/low-temperature (HFLT) corner of the SC phase of the H-T diagram~\cite{bianchi:prl-03a,radovan:nature-03}. In
addition, \Co\ is in the clean limit, with an electron mean free path on the order of a few microns within the superconducting state at low
temperature~\cite{movshovich:prl-01}. Because of these favorable conditions (FFLO was traditionally expected to be readily destroyed by impurities),
it was suggested that the additional HFLT SC phase might indeed be a realization of a long sought after FFLO state.

\begin{figure}[t]
\includegraphics[height=6.5cm,keepaspectratio]{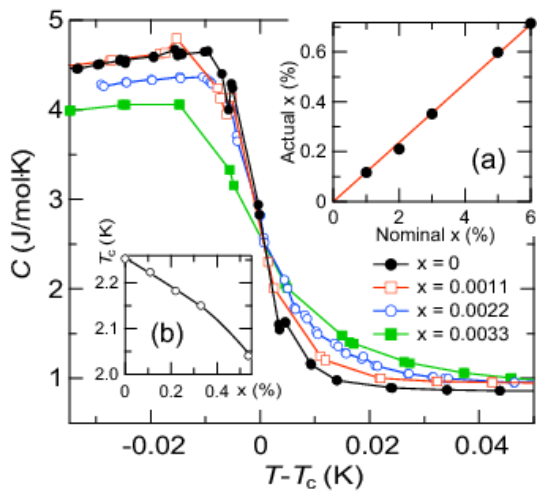}
\caption{(Color online) Specific heat of CeCo(In$_{\rm 1-x}$Cd$_{\rm x}$)$_5$ (x=0.011, 0.022 and 0.033) around $T_c$, plotted against reduced
temperature $T-T_c$. (a) Actual vs. nominal Cd concentration. Red solid line is a fit to the data through the origin. (b) $T_c$ as a function of
nominal Cd concentration x. Line is a guide to the eye.} \label{doping}
\end{figure}

There have been a number of thermodynamic, transport, and microscopic investigations of the HFLT phase in \Co\ (for a recent review see
Ref.~\cite{matsuda:jpsj-07}). Many studies have been interpreted as supporting the FFLO nature of the HFLT phase. A recent NMR
investigation~\cite{YoungBL:Micefm} concluded, however, that there is a long range antiferromagnetic order within the HFLT phase, making it, at
least, a more complicated version of an FFLO state. Regardless of whether the HFLT state is of purely magnetic origin, or magnetism accompanies a
fundamentally FFLO state, the magnetism is stabilized in the superconducting state only and does not extend into the normal state. This is a highly
unique situation, exactly opposite to the canonical picture of the competition between superconductivity and magnetism, and is worthy of detailed
experimental and theoretical studies. Additional investigations of the HFLT state in \Co\ are required before a firm case can be made for its nature.

Pressure has proven to be a very useful tuning parameter in the quest to elucidate the connection
between magnetism and superconductivity in \Co. It was shown~\cite{MicleaCF:PredFs} that pressure
enhances both $T_0$ and the extent of the HFLT state, while it suppresses the QCP~\cite{RonningF:Presqc}
suggested to arise from a nearby AFM ground state~\cite{bianchi:prl-03b}. The opposite effect of
pressure on the HFLT and AFM lead the authors to conclude that the HFLT state is of an FFLO origin~\cite{MicleaCF:PredFs}. Recently, it was shown that Cd doping in \Co\ suppresses
superconductivity and stabilizes antiferromagnetism, finally exposing the AFM state~\cite{pham:prl-06}
that might be responsible for the QCP at $H_{c2}$ in \Co. The authors also demonstrated that Cd doping
effect can be reversed by pressure, which drives the system back from AFM to the SC ground state. Since
Cd doping stabilizes AFM state, while pressure suppresses it and instead stabilizes the HFLT phase,
investigation of the effect of Cd impurities on the HFLT can provide important clues about its nature.

\begin{figure}[t]
\includegraphics[width=7.5cm,keepaspectratio]{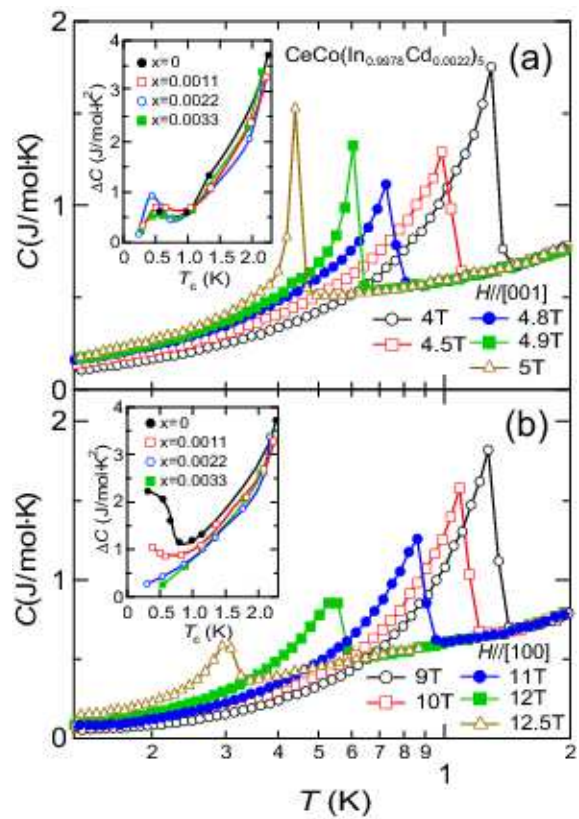}
\caption{(Color online) Electronic specific heat of CeCo(In$_{\rm 0.98}$Cd$_{\rm 0.02}$)$_5$ vs temperature for (a) $H\parallel$[001] and (b)
$H\parallel$[100]. Insets: Jump height at $T_c$ vs. $T_c(H_{c2})$. All solid lines are guide to the eye.} \label{T0}
\end{figure}

In this Letter, we present the results of specific heat measurements on CeCo(In$_{\rm 1-x}$Cd$_{\rm x}$)$_5$ with low Cd (x=0.0011, 0.0022 and
0.0033) and Hg (0.00016, 0.00032, and 0.00048) concentrations, which directly address the stability of the first order nature of the SC transition
and the HFLT phase of \Co\ with respect to impurities. Single crystals of \CoCd\ and \CoHg\ were grown from In-flux. Single plate-like samples with a
typical weights of 1-3 mg were used for specific heat measurements. Initial sample characterization via micro-probe analysis, using wavelength
dispersive spectroscopy, showed uniform distribution of the dopants. The actual Cd concentration, shown in Fig.~1(a), is linear with respect to the
nominal concentration, with zero offset. The actual/nominal concentration ratio of ~0.11 is in good agreement with ~0.1 reported by L. D. Pham {\it
et al.}~\cite{pham:prl-06}. The actual concentrations, rather than the nominal ones, are referred to in the rest of this Letter. The specific heat at
temperatures down to 100\,mK and high fields up to 14\,T  was measured in a dilution refrigerator and a superconducting magnet, employing the
quasi-adiabatic method. Physical Property Measurement System (PPMS) from Quantum Design was used to measure specific heat at low fields in the
vicinity of the SC anomaly.

The main panel of Fig.~\ref{doping} shows the zero field specific heat of CeCo(In$_{\rm 1-x}$Cd$_{\rm x}$)$_5$, with x=0.0011, 0.0022 and 0.0033, against the reduced temperature $T-T_c$ in the vicinity of the SC transition. With increasing x, the jump decreases and the width of the specific heat anomaly at SC transition increases monotonically, while $T_c$ decreases linearly (inset (b)). These monotonic changes in SC properties indicate a gradual variation of the actual Cd concentration with x, consistent with the results of the micro-probe analysis. These results show good control of the amount of the Cd dopants and their homogeneous distribution in our samples.

\begin{figure}[t]
\includegraphics[width=7.5cm,keepaspectratio]{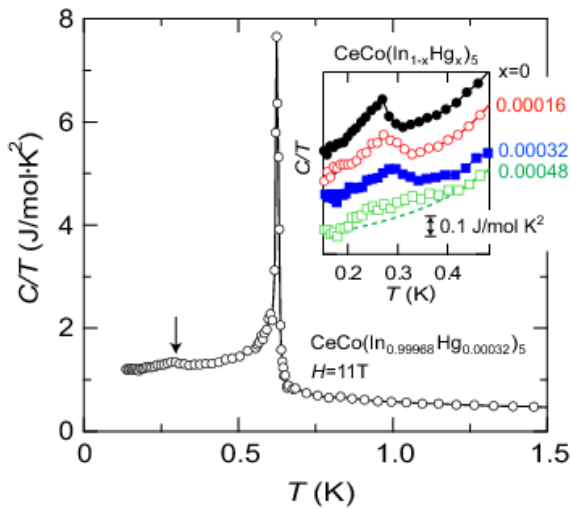}
\caption{(Color online) Electronic specific heat divided by temperature $C/T$ for CeCo(In$_{\rm 0.99968}$Hg$_{\rm 0.00032}$)$_5$ as a function of
temperature at magnetic field 11\,T applied along [100]. Arrow indicates an anomaly within the SC state. Inset: Expanded plot of $C/T$ around the
transition temperature into the HFLT SC phase. Data are shifted vertically for clarity. Dotted line is a smooth background between the high and low
temperature data.} \label{FFLO}
\end{figure}

Figure~\ref{T0} shows the effect of Cd doping on the SC anomaly in specific heat at high magnetic field. The specific heat of CeCo(In$_{\rm
0.9978}$Cd$_{\rm 0.0022}$)$_5$ at low temperatures and fields close to $H_{c2}$ is shown in the main panels after subtraction of the low temperature
tail due to a nuclear Schottky anomaly.  Fig.~\ref{T0}(a) shows the specific heat for $H\parallel$[001]. The jump in the second-order-like SC anomaly
initially decreases with increasing field, but above 4.8\,T the height of the anomaly increases, its width narrows, and the shape becomes more
symmetric, indicating the change in the order of the SC transition from second to first for $H > 4.8$ T. This is in contrast to the evolution of the
SC anomaly for $x = 0.0022$ with $H\parallel$[100], shown in the main panel of Fig.~\ref{T0}(b), where the size of the jump at $T_c$ decreases
monotonically with increasing field. These data indicate that the first order character of the SC transition at low temperatures is suppressed
already with 0.2\% Cd-doping for $H\parallel$[100]. Similar data for different levels of Cd-doping are summarized in the insets of Fig.~\ref{T0},
where we plot the jump in specific heat at $T_c$, with magnetic field as an implicit variable via $H_{c2}(T_c)$. $\Delta C$ is monotonic for $x \ge
0.02$ and $H\parallel$[100] (inset of Fig.~\ref{T0}(b)) (no first order transition), while all other curves are non-monotonic (signature of the first
order phase transition). Cd-doping is therefore more effective in suppressing the first order nature of the SC transition when the field is applied
within the $a-b$ plane of \CoCd, in spite of the fact that the height of the anomaly itself is greater in pure \Co\ for this field orientation.

There is no indication of an additional specific heat anomaly within the SC phase (as the one defining the HFLT phase in pure \Co) for the Cd-doped
samples studied. Given that the lowest Cd concentration studied is only $\approx 0.1$\%, the proposed FFLO state appears to be extremely susceptible
to impurities, in agreement with the theoretical work by Adachi, {\it et. al}~\cite{AdachiH:EffPpt}.

To probe the effect of even lower impurity concentrations, we conducted specific heat investigation of Hg-doped samples \CoHg, with Hg concentrations
of x = 0.00016, 0.00032, and 0.00048. The specific heat data for a sample with x = 0.00032 in several fields close to $H_{c2}$ are displayed in
Fig.~\ref{FFLO} after subtraction of the nuclear contribution. In addition to the first order SC anomaly, the anomaly associated with the lower
temperature phase transition into the HFLT state is clearly resolved. Such anomalies for all the Hg-doped samples are displayed in the inset to
Fig.~\ref{FFLO}. The anomaly does not shift substantially in temperature with increase in Hg doping level, but it is suppressed in an unusual way.
The temperature of the anomaly is not driven to $T = 0$, but, instead, the anomaly gradually broadens and eventually washes out.

\begin{figure}[t]
\includegraphics[width=7.5cm,keepaspectratio]{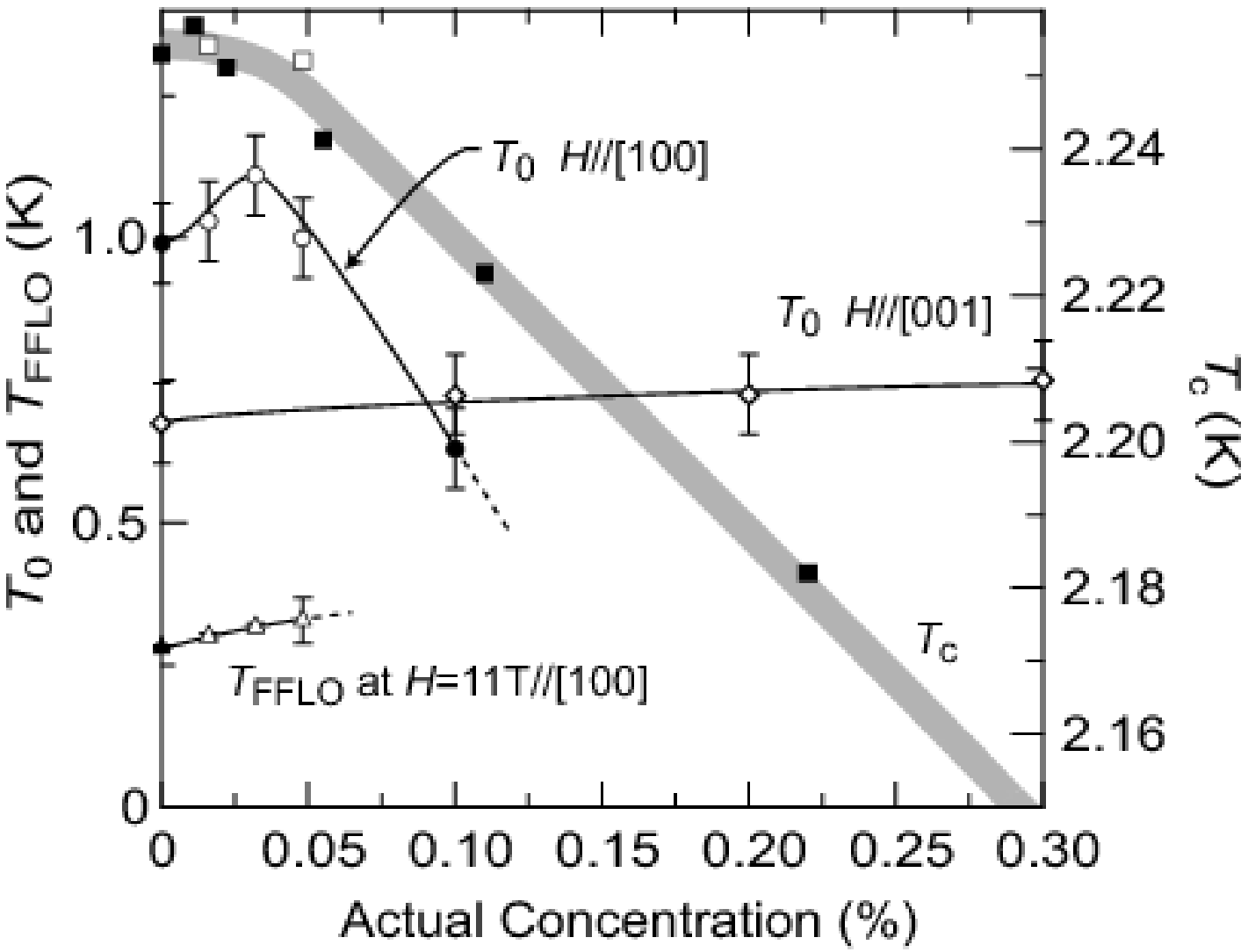}
\caption{Doping dependence of the onset of first-order phase transition $T_0$, $T_{\rm FFLO}$ at magnetic field $H$=11\,T applied along [100] and
superconducting transition temperature $T_c$ at zero field. Solid symbols: \CoCd \,\,(x=0, 0.0011, 0.0022 and 0.0033). Open symbols: \CoHg\ (x =
0.00016, 0.00032 and 0.00048). $\medbullet$,\,$\medcirc$: $T_0$ for $H\parallel$[100], $\Diamond$: $T_0$ for $H\parallel$[001], $\blacksquare$,
$\Box$: $T_c$, $\bigtriangleup$, $\blacktriangle$: $T_{\rm FFLO}$. All curves are guides to the eye.} \label{T0TffloTc}
\end{figure}

Figure~\ref{T0TffloTc} summarizes our results for both the first order SC phase transition, the HFLT anomaly, and a zero field SC transition
temperature $T_c$ as a function of both Cd and Hg dopant concentration. The narrow region where $T_c$ is independent of doping concentration is most
likely due to the break down of the homogeneous Abrikosov-Gorkov (AG) model of the impurity suppression of $T_c$, which starts out with a linear
slope of $T_c$ vs. $x$. Impurities suppress superconductivity within the region on the order of the coherence length $\xi$. When the inter-impurities
distance $d > 2 \xi$, the AG model is expected to break down (inhomogeneous SC limit), and $T_c$ should become independent of impurity concentration.
We can estimate $d = \left(V \over {5 \times 5\times 10^{-4} }\right)^{1/3} = 40$\,\AA, where $V = 161$ \AA$^3$ is the unit cell volume of
\Co~\cite{petrovic:jpcm-01} and 5$\times$10$^{-4}$ is the approximate critical doping concentration for suppressing the HFLT phase. The estimated $d$
compares well with $2 \xi \approx 70$\AA\,\,obtained by previous thermodynamic measurements~\cite{IkedaS:UncsC5}, demonstrating that $\xi$ is the
relevant length scale for the HFLT state. Therefore, the HFLT state is likely of the superconducting, non-magnetic origin.

Earlier theoretical studies of impurity effects on the first order SC transition by Maki and Tsuneto~\cite{maki:ptp-64} used microscopic theory to
show that for an s-wave SC in a strong Pauli limit, the SC transition remains first order below $T_0 = 0.56\ T_c$ for non-magnetic impurities.
Conversely, it was widely accepted that an FFLO state is easily suppressed by small amount of impurities, perhaps based on the results of the early
theoretical investigations of FFLO in s-wave superconductors~\cite{aslamazov:jetp-67}. Recent theoretical studies of impurity effects on FFLO states
in d-wave superconductors came down on both sides of the issue, some suggesting a moderately sensitive nature of FFLO state~\cite{AdachiH:EffPpt},
and some concluding that an FFLO state is robust against impurites~\cite{agterberg:jpcm-01,Vorontsov:private-07}. If the HFLT state is indeed of an
FFLO origin, as suggested by this work, our results support a very fragile nature of the FFLO state with respect to impurities.

If the HFLT state was of magnetic origin, we would expect the enhancement of such a state with Cd and Hg impurities, since higher concentrations of
$\approx 0.5$\% stabilize an AFM ground state. Instead, the HFLT is suppressed at very low concentrations, suggesting competition of the AFM and the
HFLT state. This competition may provide an avenue for suppression of the FFLO state in addition to a simple impurity scattering effect, and may be
responsible for the high sensitivity of the HFLT state to Cd impurities. This effect was not taken into account by recent theoretical
investigations~\cite{agterberg:jpcm-01,Vorontsov:private-07}, which may reconcile them with our experimental results.

In conclusion, we have measured the specific heat of CeCo(In$_{\rm 1-x}$Cd$_{\rm x}$)$_5$ with x=0.0011, 0.0022 and 0.0033 and \CoHg\ with x=0.00016,
0.00032, and 0.00048 to study the doping effect on high field-low temperature SC state of CeCoIn$_5$. We found that roughly 0.05\,\% of Hg-doping is
sufficient to suppress the HFLT state. Thus, the HFLT state is extremely sensitive to impurities, suggesting a non-magnetic, FFLO origin of this
state. The first order character of the SC transition is less susceptible to impurities, with an anisotropic response to Cd-doping. The cross-over
temperature $T_0$, where SC transition changes its character from first to second order, decreases rapidly with increasing Cd-doping for
$H\parallel$[100], while it remains roughly the same up to x=0.003 for $H\parallel$[001]. The relative robustness to impurities of the first order
transition compared to the HFLT state is in agreement with the theoretical calculations of Ref~\cite{AdachiH:EffPpt} for the FFLO state.

We are grateful to I Vekhter, L. Boulaevskii, M. Graf, J. Sauls, and A. Balatsky for stimulating discussions. Work at Los Alamos National Laboratory
was performed under the auspices of the U.S. Department of Energy.  A. D. Bianchi received support from NSERC (Canada), FQRNT (Qu\`{e}bec), and the
Canada Research Chair Foundation. Z. Fisk acknowledges support NSF grant NSF-DMR-0600742. We thank Sarah Roeske in the UC-Davis Geology Department
for assistance with the microprobe experiments. Work at UC-Davis was supported by the NSF grant No. DMR-0600742.

\end{document}